\documentclass[debug]{rmaa}
\usepackage{paralist}
\usepackage{psfrag,color}
\usepackage[latin1]{inputenc}

\title{SNR radio spectral index distribution and its correlation 
with polarization: a case study of Lupus Loop} 

\author{
  V. Borka Jovanovi\'{c},\altaffilmark{1} 
  P. Jovanovi\'{c},\altaffilmark{2}
  and D. Borka\altaffilmark{1}}
 
\altaffiltext{1}{Vin\v{c}a Institute of Nuclear Sciences, University 
of Belgrade, Serbia.}

\altaffiltext{2}{Astronomical Observatory, Belgrade, Serbia.}

\shortauthor{Borka Jovanovi\'{c}, Jovanovi\'{c}, \& Borka}
\shorttitle{Radio spectral index of Lupus Loop}

\fulladdresses{
\item Vesna Borka Jovanovi\'{c} and Du\v{s}ko Borka: Atomic Physics 
Laboratory (040), Vin\v{c}a Institute of Nuclear Sciences, University 
of Belgrade, P.O. Box 522, 11001 Belgrade, Serbia.
\item Predrag Jovanovi\'{c}: Astronomical Observatory, Volgina 7, 
P.O. Box 74, 11060 Belgrade, Serbia.}

\listofauthors{V. Borka Jovanovi\'{c}, P. Jovanovi\'{c}, \& D. Borka}
\indexauthor{Borka Jovanovi\'{c}, V.}
\indexauthor{Jovanovi\'{c}, P.}
\indexauthor{Borka, D.}

\abstract{We use radio-continuum all-sky surveys at 1420 and 408 MHz 
with the aim to investigate properties of the Galactic radio source 
Lupus Loop. The survey data at 1435 MHz, with the linear 
polarization of the southern sky, is also used. We calculate 
properties of this supernova remnant: the brightness temperature, 
surface brightness and radio spectral index. For determining borders 
and calculation of its properties, we use the method we have 
developed. The non-thermal nature of its radiation is confirmed. The 
distribution of spectral index over its area is also given. A 
significant correlation between the radio spectral index distribution 
and the corresponding polarized intensity distribution inside the 
loop borders is found, indicating that the polarization maps could 
provide us information about the distribution of interstellar medium, 
and thus could represent one additional way to search for new 
Galactic loops.}

\resumen{}

\addkeyword{ISM: supernova remnants}
\addkeyword{radiation mechanisms: non-thermal}
\addkeyword{radio continuum: ISM}

\begin{document}
% Typeset article header
\maketitle

\section{Introduction}

Radio surveys of the region in vicinity of the supernova of 1006 
A.D. revealed a plateau or spur running out from the galactic plane 
near $l$ = 330$^\circ$ which contains two shell-like objects: Lupus 
Loop and SN1006 \citep{miln71}. The general diffuse appearance of 
the Lupus Loop suggested that it is the remnant of a very old 
supernova (which is consistent with its low surface brightness) and 
these two remnants are not associated in any way \citep{miln74}. It 
was indicated earlier \citep{spoe73} that expanding sphere of the 
supernova remnants (SNRs) leads to compression of the interstellar 
magnetic field, which results in an observable radio source, and 
that the spatial orientation of the loops contains information on 
the direction of the magnetic field of the undisturbed medium 
outside the shell. \citet{spoe73} treated the Lupus Loop as an 
object similar to the main Galactic loops, and found that it 
indicates magnetic field direction parallel to field found from Loop 
I. Radio continuum observations of this source are given in 
\citet{miln71,miln74}, radio line observations in \citet{colo82}, 
$X$-ray observations and its spectrum can be found in 
\citet{toor80,leah91,ozak94,kapl06} and references therein, while 
far $UV$ observations are presented in \citet{shin06}.

The analysis on the filamentary structure observed in polarization 
by WMAP (Wilkinson Microwave Anisotropy Probe) satellite is given in 
\citet{vida15,vida16}. It is described there that most of the 
polarized emission (at high latitudes) comes from individual 
filamentary features, and some of these structures are the 
well-known continuum radio loops. Using WMAP data at 23, 33 and 41 
GHz, they studied the diffuse polarized emission over the entire 
sky, and they obtained the (average) polarization spectral indices 
which are consistent with synchrotron radiation.

A catalogue of Galactic SNRs, with some statistics of their 
parameters, is presented in \citet{gree14a}, along with more 
detailed web-based version \citep{gree14b}. The current version of 
the catalogue contains 294 SNRs, and is based on research in the 
published literature up to the end of 2013. Lupus Loop, with 
catalogue name G330.0+15.0, is listed there, and some of its 
parameters are given. This low surface brightness loop has been 
observed in radio and $X$-ray wavelength range \citep{gree14b}.

Our aim is to study the properties of this remnant, and to calculate 
radio spectral index using the method we have previously developed. 
Our method of calculation is explained in detail in \citet{bork12b} 
and references therein: we investigated Galactic Loops I-VI in 
papers \citet{bork06a,bork06b,bork07},
\newline \citet{bork08a,bork10,uros11}, and we investigated smaller 
remnants in \citet{bork08b,bork09a,bork09b,bork10,bork11},
\newline \citet{bork12a}. This method is applicable to extragalactic 
radio sources as well \citep{bork12c}. In this paper we also want to 
investigate the nature of its radiation and to study how spectral 
index varies across the face of the remnant. Besides, we want to 
analyze and study the polarization of this SNR and to investigate 
its connection with the spectral index.

\begin{figure}
\centering
\includegraphics[width=0.45\textwidth]{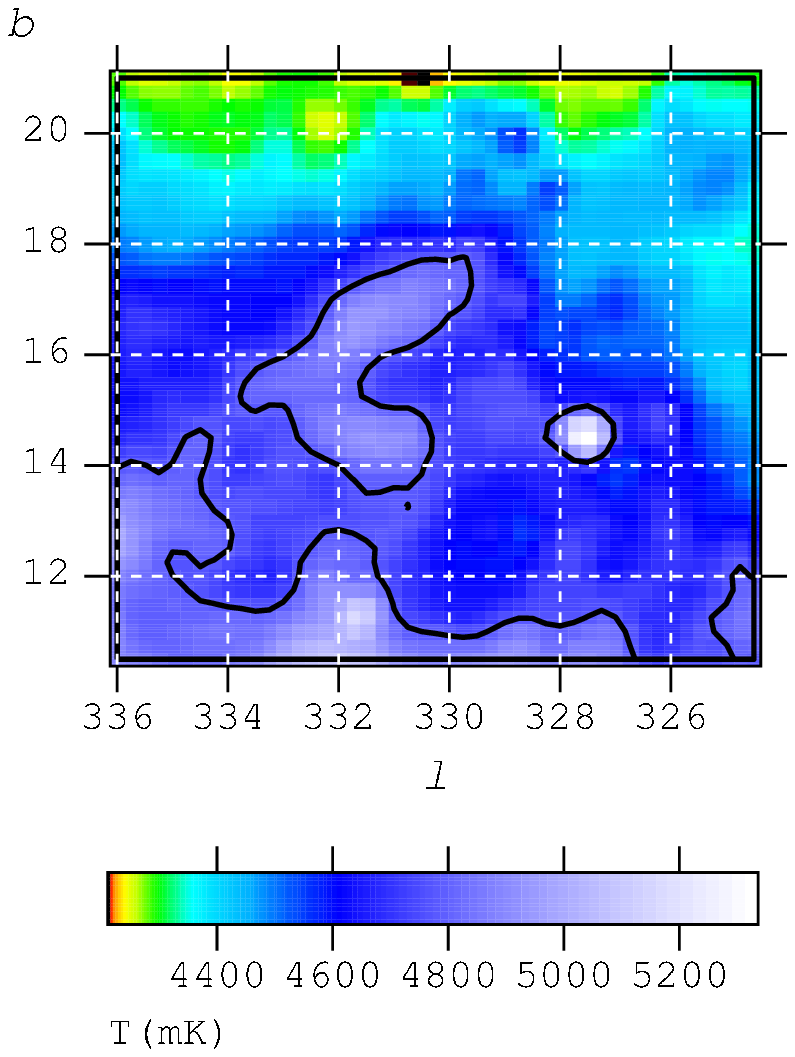}
\hspace{0.5cm}
\includegraphics[width=0.45\textwidth]{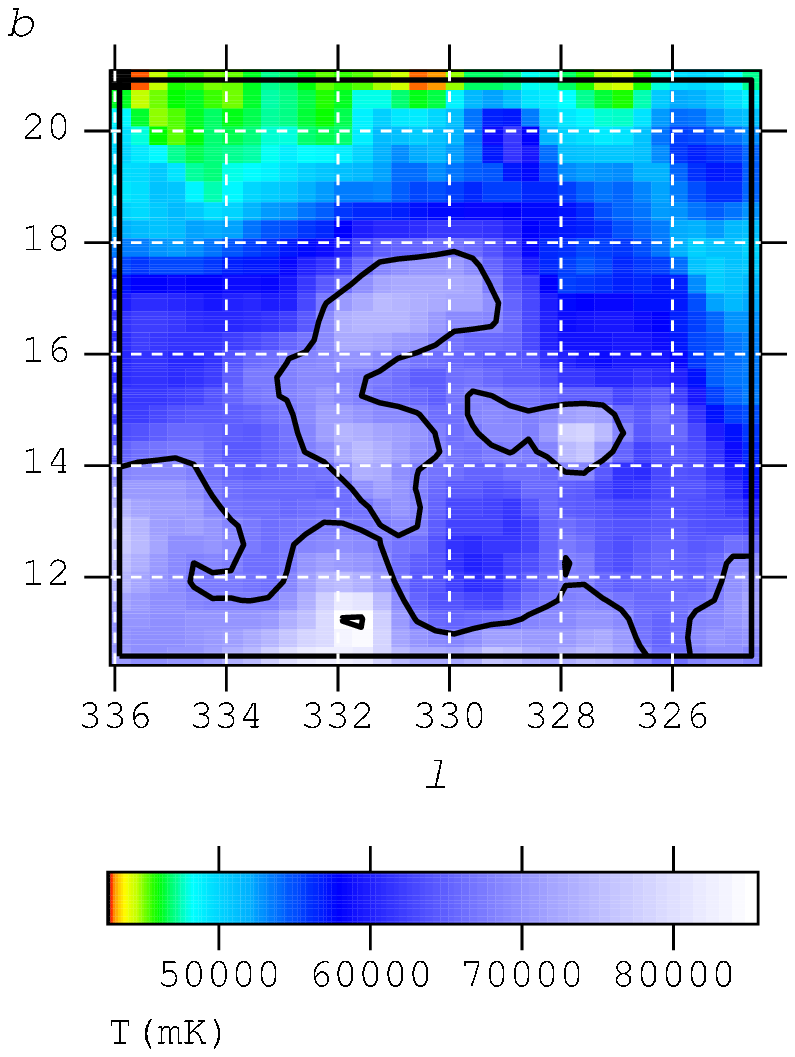}
\caption{Lupus Loop area with brightness temperature contour at 
1420 MHz \textbf{(left)} and 408 MHz \textbf{(right)}. The contour 
is representing $T_{min}$ as given in the Table 1. Below, the 
temperature scales are given (in mK).}
\label{fig01}
\end{figure}

\begin{figure}
\centering
\includegraphics[width=0.95\textwidth]{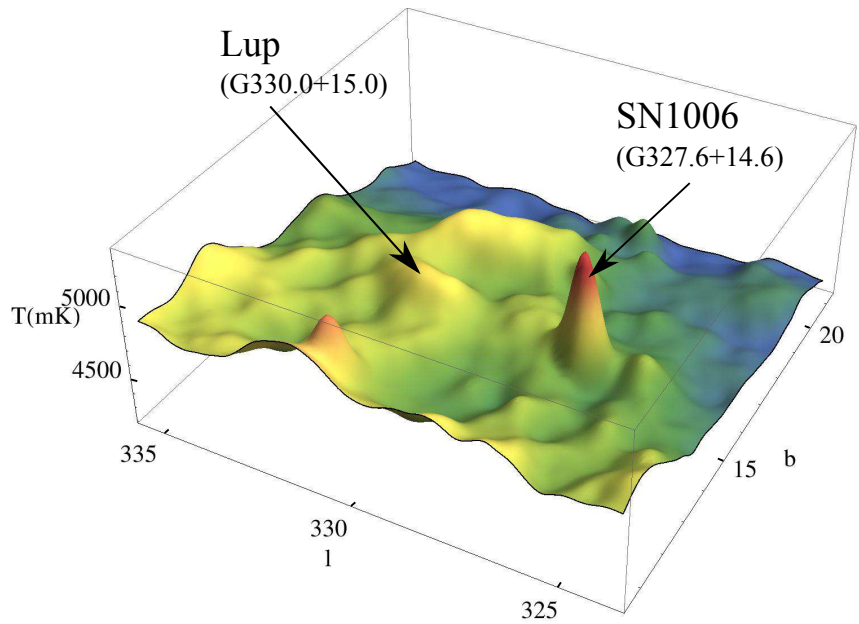}
\caption{The 3D map of the Lupus Loop and its surrounding at 1420 
MHz. The brightness temperature is given in mK.}
\label{fig02}
\end{figure}

\begin{figure}
\centering
\includegraphics[width=0.48\textwidth]{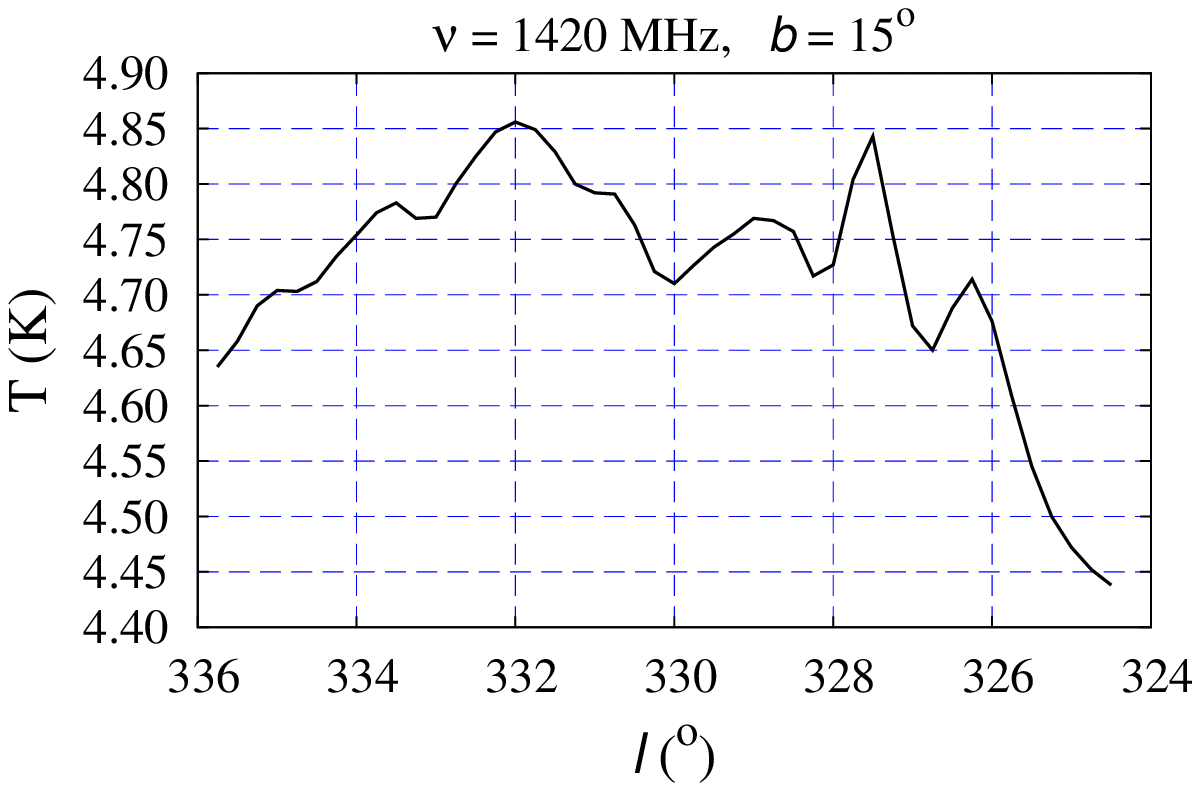}
\hspace{0.2cm}
\vspace{0.5cm}
\includegraphics[width=0.48\textwidth]{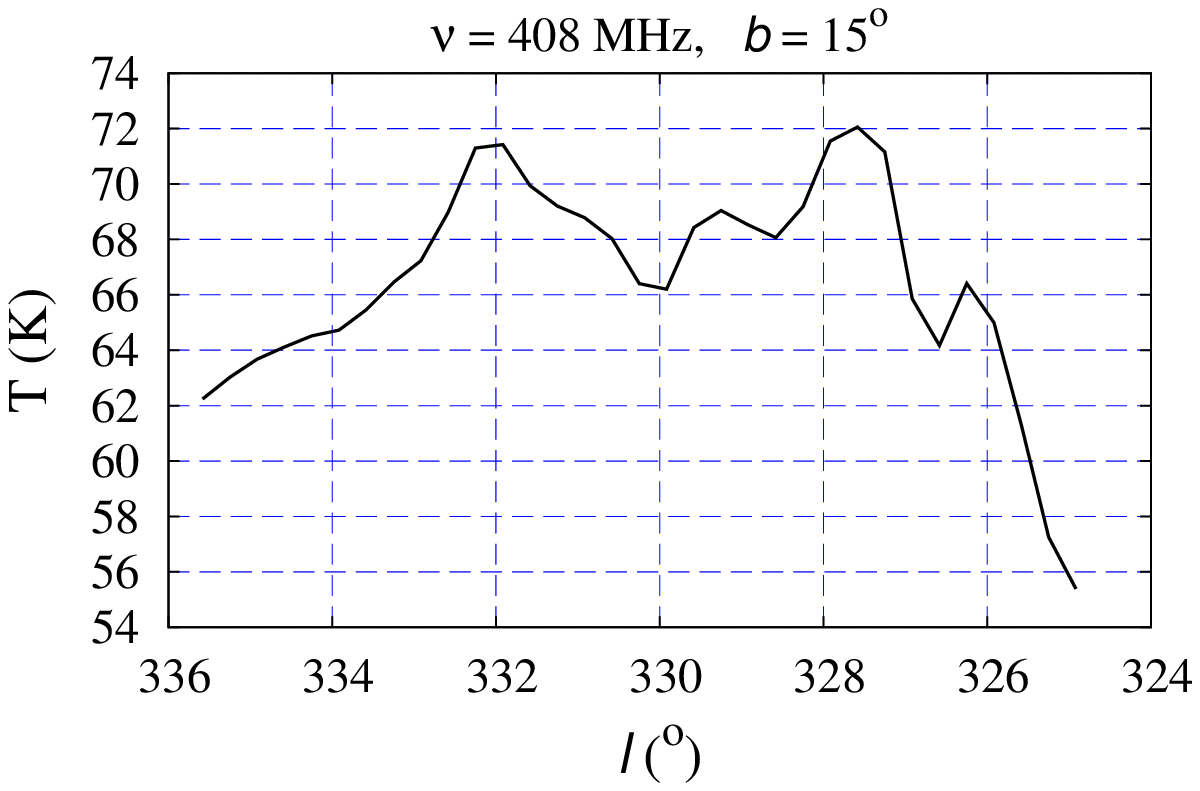}
\includegraphics[width=0.48\textwidth]{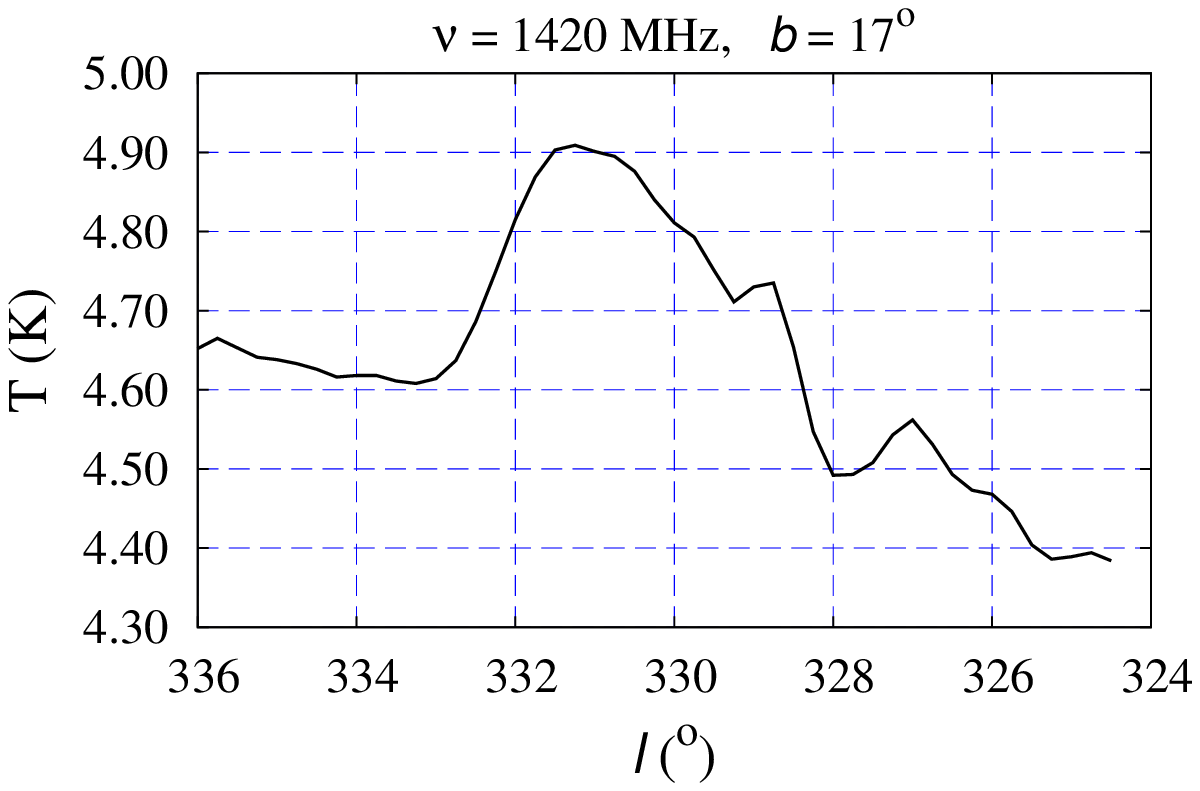}
\hspace{0.2cm}
\includegraphics[width=0.48\textwidth]{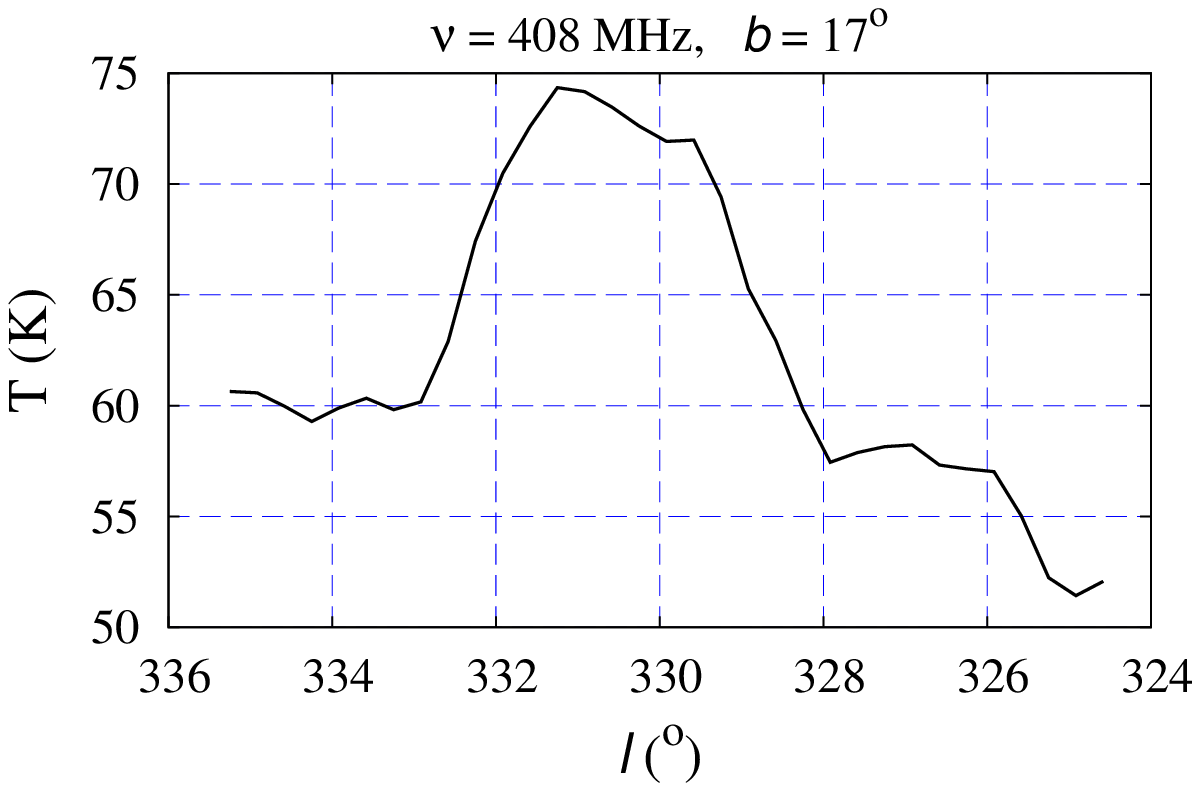}
\caption{\textbf{Left:} 1420 MHz temperature profiles for area 
containing Lupus Loop at galactic longitude $b$ = 15$^\circ$ (top) 
and 17$^\circ$ (bottom). \textbf{Right:} 408 MHz temperature 
profiles for $b$ = 15$^\circ$ (top) and 17$^\circ$ (bottom).}
\label{fig03}
\end{figure}

\section{Lupus Loop area}

\subsection{Temperature brightness contours}

We use the radio continuum surveys of the sky provided by Max Planck 
Institute for Radio Astronomy (MPIfR), Bon, Germany, available at 
the internet site: \url{http://www3.mpifr-bonn.mpg.de/survey.html}. 
Observational data, which we use for our calculations, are obtained 
from continuum radio emission at: 1420 MHz \citep{reic01} and 408 
MHz \citep{hasl82}. At the frequency of 34.5 MHz \citep{dwar90} the 
loop could not be resolved. The angular resolutions of the surveys 
are: 35' at 1420 MHz and 0$^{\circ}$.85 at 408 MHz, which give the 
corresponding observations at the rates, for both $l$ and $b$, 
(1/4)$^{\circ}$ (1420 MHz) and (1/3)$^{\circ}$ (408 MHz). The 
effective sensitivities for average brightness temperature ($T_b$) 
are about 50 mK (1420 MHz) and 1.0 K (408 MHz).

As we showed earlier (\citet{bork12b} and references therein), the 
method for defining a loop border and for determining the values of 
temperature and brightness, which we developed for main Galactic 
Loops I-VI, could be applicable to all SNRs. In that way we have
determined the Lupus Loop area which is shown in Fig. \ref{fig01}. 
There is also influence of some other radio sources in the 
vicinity of the Lupus, regarding the determination of its radio 
properties, like brightness and radio spectral index. The 3D plot 
showing brightness temperatures of this loop and its surrounding, at 
the frequency of 1420 MHz, we show in Fig. \ref{fig02}. To make this 
figure more clear, we denoted the ridges of Lup and SN1006 with 
arrows. In Fig. \ref{fig03} we want to demonstrate a few 
temperature profiles. Note that temperature peak around $l$ = 
328$^\circ$ corresponds to another source (SN1006), and that we have 
to pay attention to the part of the profiles from $l$ = 334$^\circ$ 
to 329$^\circ$ to see how it changes over the Lupus SNR.

\subsection{Brightness temperatures and surface brightnesses}

Minimum and maximum brightness temperatures are given in Table 
\ref{tab01}. We use these values to determine the loop borders, and 
then, after subtracting the background emission over the Lupus area, 
we determine its brightness temperature $T_b$. Assuming the spectra 
to have a power-law form, i.e. the flux density to be proportional 
to the frequency $S_{\nu} \propto \nu^{-\alpha}$ (or $T_b \propto 
\nu^{-\beta}$), using data of at least two frequencies, we can 
calculate radio spectral index $\alpha$ (or $\beta = \alpha + 2$) by 
fitting this equation to the data. Knowing two values of 
brightnesses (derived in this paper, at 1420 and 408 MHz), we 
obtained radio spectral index $\alpha$ = 0.98. Using the relation:

\begin{equation}
\Sigma_{\nu} = 2kT_{b,\nu} (\nu/c)^2,
\end{equation}

\noindent where $k$ is the Boltzmann constant and $c$ the speed of 
light, we calculate the surface brightness $\Sigma_{\nu}$, and with

\begin{equation}
\Sigma_{1000} / \Sigma_{\nu} = (1000/\nu)^{-\alpha}
\end{equation}

\noindent we finally obtain the values reduced at 1000 MHz. The 
results are listed in Table \ref{tab01}.

\begin{table}[!t]
\centering
\setlength{\tabnotewidth}{0.5\columnwidth}
\tablecols{6}
% Stretch the space between table columns 
\setlength{\tabcolsep}{0.5\tabcolsep}
\caption{Temperature borders, temperatures and brightnesses of the 
Lupus Loop at 1420 and 408 MHz. The brightnesses are reduced to 1000 
MHz with the spectral index we calculated: $\alpha$ = 0.98.}
\begin{tabular}{rccccc}
\toprule
frequency & $T_{min}$ & $T_{max}$ & temperature & brightness  
& brightness reduced\\
(MHz) & (K) & (K) & (K) & (10$^{-23}$ W/(m$^2$ Hz sr)) & 
to 1000 MHz \\
& & & & & (10$^{-23}$ W/(m$^2$ Hz sr)) \\
\midrule
\textbf{1420} & 4.78 & 5.4 & 0.10 $\pm$ 0.05 & 0.63 
$\pm$ 0.31 & 8.82 $\pm$ 4.36 \\
\textbf{408} & 67.8 & 85 & 4.1 $\pm$ 1.0 & 2.12 
$\pm$ 0.51 & 8.82 $\pm$ 2.13 \\
\bottomrule
\end{tabular}
\label{tab01}
\end{table}

\section{Radio spectral index}

\begin{figure}[ht!]
\centering
\includegraphics[width=0.48\textwidth]{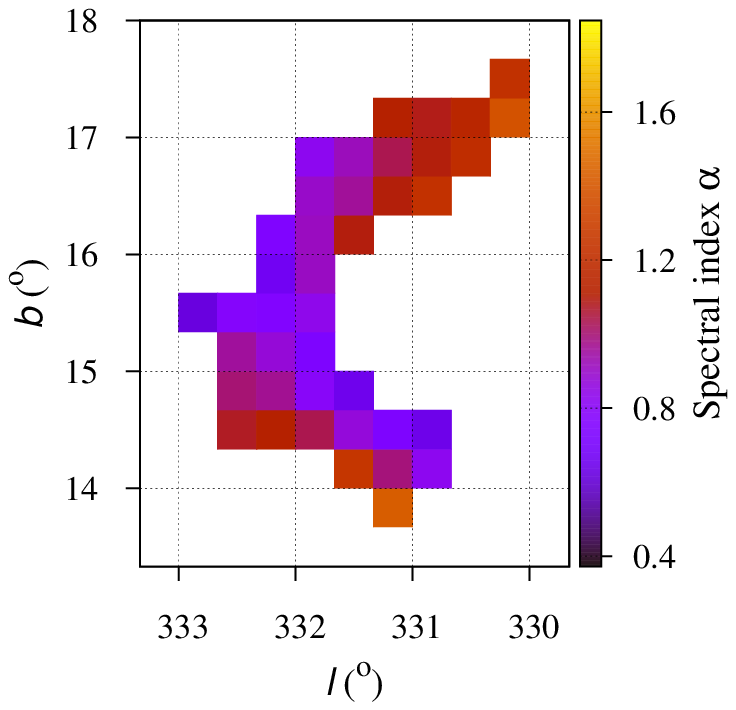}
\hfill
\includegraphics[width=0.5\textwidth]{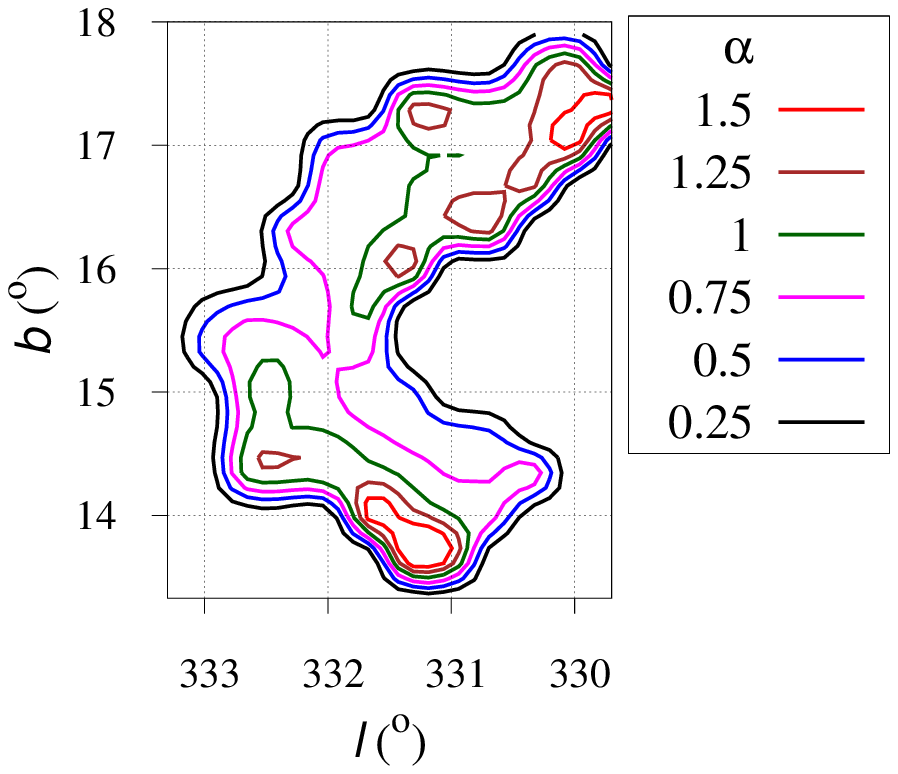}
\caption{Radio spectral index distribution across the Lupus Loop 
between 1420 and 408 MHz, in form of color map 
\textbf{(\textit{left})} and interpolated contours 
\textbf{(\textit{right})}.}
\label{fig04}
\end{figure}

We calculated the mean value of the radio spectral index between 
1420 and 408 MHz. With $\alpha$ = 0.98, we confirmed non-thermal 
emission of radiation for this remnant. Our motivation is also to 
study how spectral index varies across its area. The distribution of 
radio spectral indices, over the Lupus Loop area, is shown in Fig. 
\ref{fig04}.

If we compare our value for $\alpha$ with earlier results, these 
new observations yielded a greater value. \citet{miln71} 
calculated spectral index between the following frequencies: 5000, 
2700, 1614, 1410, 635, 408 and 160 MHz, and obtained mean value 
$\alpha$ = 0.38, while \citet{miln74} used data at 2700, 1660 and 
1410 MHz which resulted in $\alpha$ = 0.5, but we have to stress 
that they mentioned that conclusions about Lupus Loop were 
uncertain and that more data were required. Our result is larger 
than the typical value for Galactic SNRs, but in Green's catalogue 
regarding spectral index about 0.5 it is mentioned that it is not 
precisely determined, and there is also a question mark as a 
notice that it should be recalculated. Previous authors probably
took into account wider area for Lupus Loop (i.e. loop together with 
one part of the background which is outside the border) and in that 
way they lowered the brightness temperature, as well as the mean 
spectral index. Also, there is a noticeable tendency for more 
recent observations to give higher values of $\alpha$ than the 
previous.

Averaged values for measured spectral indices of all-sky Galactic 
radiation vary with frequencies. For example, $\alpha = 0.55$ 
between 45 and 408 MHz \citep{guzm11}, $\alpha = 0.71$ between 408 
and 3200 MHz \citep{plat03} and $\alpha = 1.01$ between 2300 and 
3300 MHz \citep{davi06}. Spectral distribution of the remnant gives 
information about the distribution of energy of the relativistic 
electrons that produce the emission observed at given 
radio-frequencies. Like in the previously mentioned case of Galaxy 
radiation, averaged value of $\alpha$ depends on the frequencies. In 
that way we can also explain greater absolute value for $\alpha$ 
in our paper than in older results.

The very end of the Lupus Loop is mixed with strong external 
sources and it is very hard to resolve loop from the background in 
this region. If we take whole area of the loop we will get $\alpha$ 
= 0.98 and so high value of index is probably  because of influence 
of these additional strong radio sources, and the ridges of the 
loop. The largest part of Lupus area has spectral indices between 
0.4 and 0.8. If we take the area of the loop without the area 
enclosed by $\alpha = 1$ line (see green line in Fig. \ref{fig04}), 
i.e. between the contours 0.4 and 1, we get the average spectral 
index $\alpha = 0.77$, which is much closer to the typical value 
for SNRs ($\alpha = 0.5$). This area of the loop is relatively 
clean from external sources and it indicates that the mean spectral 
index is highly influenced by the loop's ridges also. The spectral 
index variations must reflect differences in the acceleration process 
from place to place within the SNR. This indicates that the increased 
emissivity could be the result of particle acceleration in the SNR 
shock (see \citet{zhan97}). Although the spectral index 
determinations based on several frequencies should be more accurate 
than these based on two frequencies, we improved method for 
extracting the background, and this is very important in these cases 
when there are confusing sources.

As it can be seen from Fig. \ref{fig04}, the distribution of radio 
spectral index tells us that its variation over the loop area is 
rather large. So, we think that distribution of spectral index over 
the loop is more adequate in description of the loop than the mean 
spectral index.

From the Fig. \ref{fig04} it can also be seen that the largest part 
of the loop's area has radio spectral index between 0.4 and 0.8. 
Then, the greater values are connected with its ridges, and as 
a whole it gives the mean value 0.98.

\section{The polarization surveys and maps}

\begin{figure}[ht!]
\centering
\includegraphics[width=0.75\textwidth]{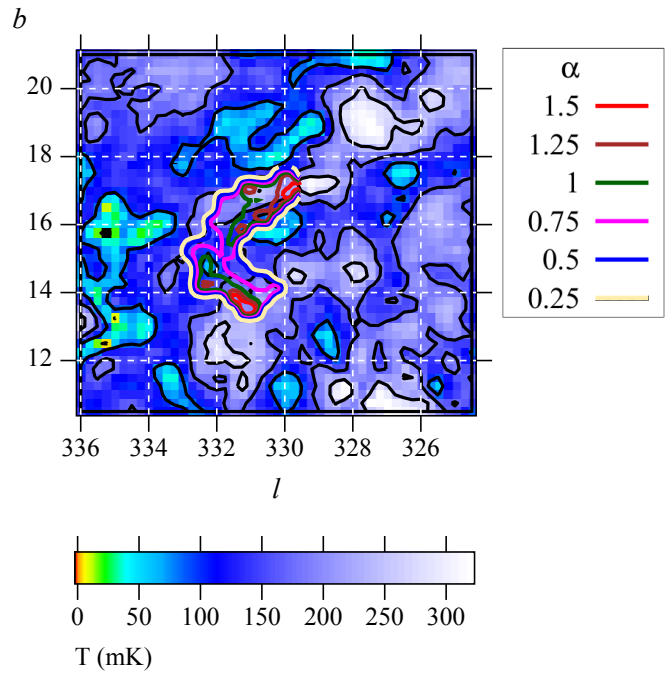}
\caption{Polarized intensity at 1435 MHz for the area surrounding 
Lupus Loop, shown in Galactic coordinates. Total intensity contours 
start at 10 mK $T_b$ and run in steps of 80 mK $T_b$. The contour 
levels of $\alpha$ from Fig. \ref{fig04} are superimposed.}
\label{fig05}
\end{figure}

\begin{figure}[ht!]
\centering
\includegraphics[width=0.46\textwidth]{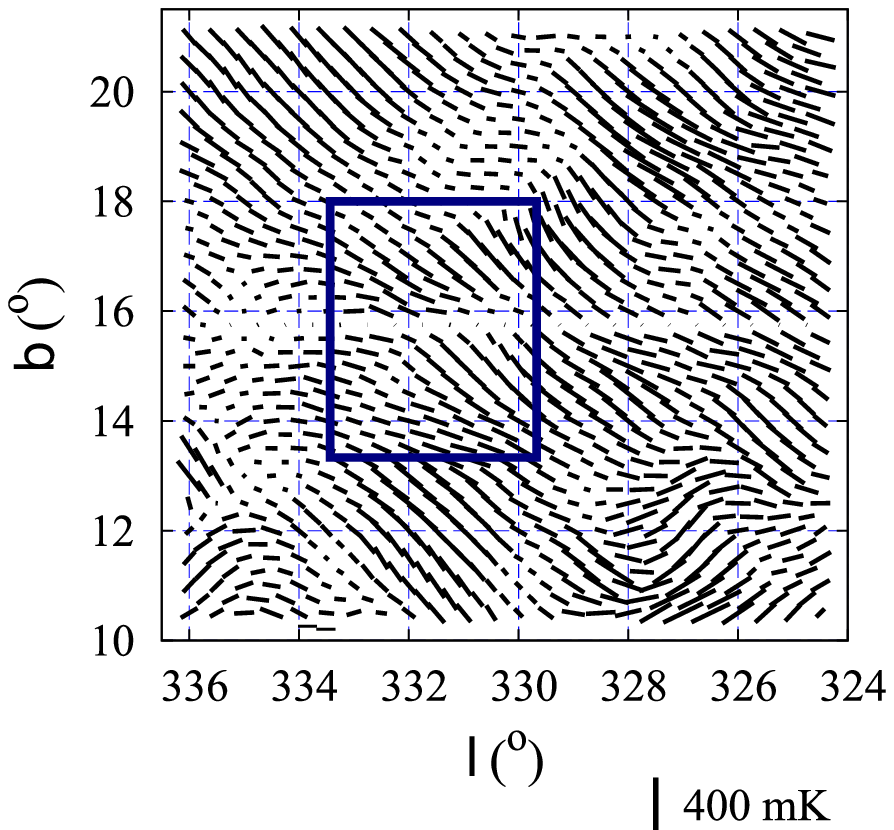}
\hspace{0.5cm}
\includegraphics[width=0.48\textwidth]{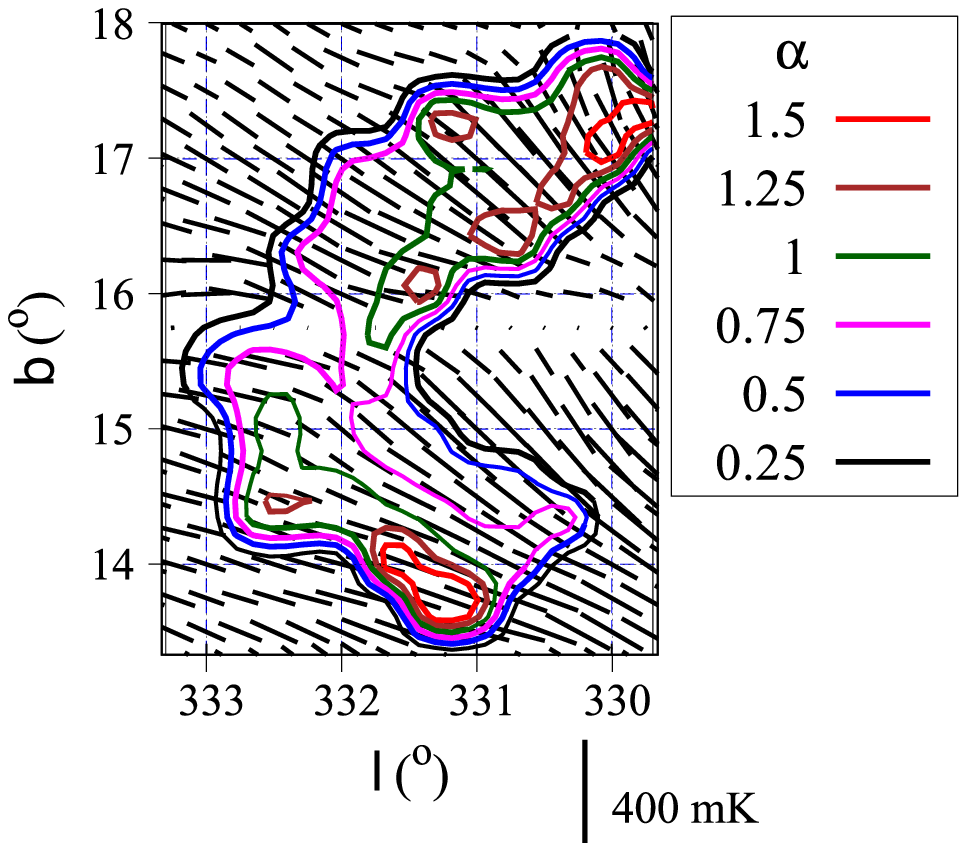}
\caption{\textbf{(\textit{Left}:)} The distribution of the 
polarization $\vec E$ vectors at 1435 MHz over the same Lupus Loop 
area as in Fig. \ref{fig01}. The length of the vectors is 
proportional to the polarized intensities (see scale at the bottom). 
The inserted rectangular labels the ($l$,$b$) area which corresponds 
to Fig. \ref{fig04}. \textbf{(\textit{Right}:)} A zoomed part of the 
figure, with the ($l$,$b$) intervals corresponding to Fig. 
\ref{fig04}. The contour levels of $\alpha$ from Fig. \ref{fig04} are 
superimposed.}
\label{fig06}
\end{figure}

\begin{figure}[ht!]
\centering
\includegraphics[width=0.65\textwidth]{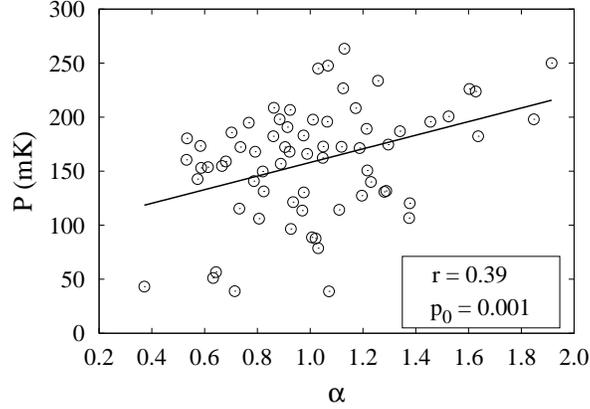}
\caption{Polarized intensity $P$ at 1435 MHz versus spectral index 
$\alpha$ between 1420 and 408 MHz, calculated at the common 
positions $(l,b)$ over the area of Lupus Loop, bounded by $T_{min}$ 
and $T_{max}$. Open circles represent $(\alpha,P)$ pairs, and 
straight solid line their fit by linear function: $P(\alpha) = 
k\alpha + \bar{P}$. The values of correlation coefficient $r$ and 
the significance level $p_0$ are denoted in the lower right corner 
of the figure.}
\label{fig07}
\end{figure}

If we want to measure the properties of the synchrotron radiation, 
it is very useful if we have the polarization data available, 
because the polarization of the mechanisms other than synchrotron is 
much smaller (see the comparison for the polarization of different 
mechanisms of radiation in \citet{vida15}). Polarized 
electromagnetic radiation can be described using the Stokes 
parameters. We can use electric field (electromagnetic wave 
propagation) $\vec E$ and define Stokes parameters $(I, Q, U, V)$ 
as time averages of the field \citep{vida15}. Parameter $I$ 
represents the total intensity of the field, parameters $Q$ and $U$ 
represent the linear polarization, while $V$ is the circular 
polarization. The polarized intensity is defined as:

\begin{equation}
P = \sqrt{Q^2+U^2},
\end{equation}

\noindent with $Q$, $U$ - the second and third Stokes parameters, 
which can be expressed by the polarization angle $\chi$ in the 
following form:

\begin{equation}
Q = P\cos{2\chi},\ \ U = P\sin{2\chi}.
\end{equation}

\noindent The degree of polarization is given as:

\begin{equation}
p = S_p/(S_p + S_u)
\end{equation}

\noindent where $S_p$ is the flux density of the polarized 
component, and $S_u$ of the unpolarized one.

The Villa Elisa survey data at 1435 MHz \citep{test08}, for linear 
polarization of the southern sky, are accessible via the MPIfR 
Survey Sampler. This linear polarization of Galactic synchrotron 
emission is given with the angular resolution of the survey of 35', 
that is, the observations are at the rates (1/4)$^{\circ}$ (for both 
$l$ and $b$). The 1435 MHz polarized intensity for the area 
surrounding Lupus Loop, we show in Fig. \ref{fig05}. From the 
polarization survey in this figure we can see how the intensity is 
distributed, and we also have appended intensity contours. Besides, 
in order to make figure more clear, we presented the superimposing 
contour levels of $\alpha$ extracted from Fig. \ref{fig04}.

For comparison of the radio continuum survey with the observations 
of polarized emission, it is very useful if we present polarized 
intensities and the polarization angles (see Fig. \ref{fig06}): the 
polarized intensities are presented by the length of the vectors, 
and the polarization angles by their orientation.

\section{Discussion}

Regarding the Figures \ref{fig01} and \ref{fig06} we can conclude 
that the polarization intensity is strongly correlated with 
brightness temperature, especially in the cases of 1420 MHz 
brightness temperature and the polarized intensity at 1435 MHz. 
Polarized emission is dominated by synchrotron radiation and because 
frequencies 1420 and 1435 MHz are very near each other, this 
correlation is expected. Polarization maps give us information about 
the interstellar distribution and it will also be one additional way 
to search for new Galactic loops.

Explanation of the large-scale polarization pattern is achieved 
using the model proposed by \citet{heil98}. In this model, an 
expanded shell compresses the magnetic field in the local 
interstellar medium (ISM). Our results indicate that the magnetic 
field should be compressed by the supernova shock. The magnetic 
field of Lupus Loop is reflected magnetic field in the ambient 
interstellar medium \citep{heil98,plat03,davi06,guzm11}.

Dependence of polarized intensity $P$ at 1435 MHz versus spectral 
index $\alpha$ between 1420 and 408 MHz over the area of Lupus Loop 
is presented in Fig. \ref{fig07}, suggesting a certain linear 
correlation between these two quantities. In order to estimate the 
significance of this linear dependence, we calculated the 
correlation coefficient $r$ and the significance level $p_0$ at 
which the null hypothesis of zero correlation is disproved. 
The relatively high value obtained for $r$ ($r = 0.39$) and small 
value for $p_0$ ($p_0 = 0.001$) both indicate that there is a
significant linear correlation between polarized intensity $P$ and 
spectral index $\alpha$ over the area of Lupus Loop as it can be 
also noticed from Fig. \ref{fig06}. We also tested linear dependence 
between these two quantities by fitting their common values with the 
function $P(\alpha) = k\alpha + \bar{P}$ (solid straight line in 
Fig. \ref{fig07}). This fit resulted in the following values: $k = 
63.0 \pm 18.4$ and $\bar{P} = 95.1 \pm 19.5$ (i.e. $\bar{P} \approx 
100 \bar{\alpha}$, where $\bar{\alpha}$ is the previously determined 
mean spectral index over the area of Lupus Loop). 

We have confirmed previous theory \citep{spoe73} that the spatial 
orientation of the loops contains information on the direction of 
the magnetic field of the undisturbed medium outside the shell.

\section{Conclusions}

As we showed earlier (\citet{bork12b} and references therein), the 
method for defining a loop border and for determining the values of 
brightness temperature and surface brightness, which we developed 
for main Galactic Loops I-VI, could be applicable to all SNRs. Here 
we use this method in order to:
\begin{itemize}
 \item determine brightness temperature borders of the Lupus Loop at 
1420 and 408 MHz,
 \item calculate the mean radio spectral index between the specified 
frequencies, as well as the distribution of indices across the face 
of this remnant,
 \item study the correlation between the radio spectral index 
distribution and the corresponding polarized intensity distribution 
within the given borders.
\end{itemize}

In the frequency range under consideration synchrotron radiation 
dominates the spectrum. We used the radio spectral index to study 
the radiation mechanism of this radio source. The value obtained for 
the spectral index (which is $>$ 0.1) confirmed non-thermal emission 
of radiation for this source. The main disagreement in the measured 
values can probably be caused by differences in the chosen area for 
Lupus Loop border. These new observations yielded value of $\alpha$ 
greater than \citet{miln74}. Besides the nature of the radiation, we 
also showed how spectral index varies across the face of the remnant.

Taking into account that SNRs radiate non-thermal (synchrotron) 
radiation which is mainly caused by the magnetic field, which on the 
other hand is also responsible for the polarization of radiation, we 
supposed that there exists the connection between the polarization 
and radio spectral index $\alpha$, which we then showed. 

We can conclude that spectral index significantly varies across the 
Lupus Loop. Over the time, the synchrotron spectral index becomes 
steeper (gets greater value, i.e. loops steepen as they age). The 
boundary of the Lupus Loop is not well defined and the ISM is rather 
inhomogeneous. That is why there are significant variations in 
spectral indices $\alpha$ over the loop area.

\textbf{Acknowledgments.} This research is part of the project 
176003 ''Gravitation and the large scale structure of the Universe'' 
supported by the Ministry of Education, Science and Technological 
Development of the Republic of Serbia.

\end{document}